# Magnetosynthesis effect on the structure and ground state of $Cu^{2+}$-based antiferromagnets

*Micaela E. Primer,*[a,#,†] *Anna A. Berseneva,*[a,#,]* *Ayesha Ulde,*[b] *Wenhao Sun,*[b] *Rebecca W. Smaha* [a,]*

[a] Materials Science Center, National Renewable Energy Laboratory, Golden, Colorado, USA

[b] Department of Materials Science and Engineering, University of Michigan, Ann Arbor, Michigan, USA

ABSTRACT

Subtle synthetic variables can have an outsizes influence on the crystal structure and magnetic properties of a material, particularly those of quantum materials. In this work, we investigate the impact of synthesis under a magnetic field (magnetosynthesis) on the crystal structure and magnetic properties of several $Cu^{2+}$ ($S$=1/2) based materials with antiferromagnetic interactions and varying levels of magnetic frustration, from simple antiferromagnets to a quantum spin liquid. We employ small (0.09 – 0.37 T) magnetic fields applied during low-temperature hydrothermal or evaporative synthesis of the simple antiferromagnet $CuCl_2·2H_2O$, the canted antiferromagnet $(Cu,Zn)_3Cl_4(OH)_2·2H_2O$, the frustrated and canted antiferromagnet atacamite $Cu_2(OH)_3Cl$, and the highly frustrated quantum spin liquid herbertsmithite $Cu_3Zn(OH)_6Cl_2$. We found that $(Cu,Zn)_3Cl_4(OH)_2·2H_2O$ experiences structural changes well above its magnetic transition.

Atacamite $Cu_2(OH)_3Cl$ synthesized under a 0.19 T field experiences a 0.15 K (~3%) decrease in its Néel transition temperature and a significant strengthening of its antiferromagnetic interactions, suggesting that magnetosynthesis can influence the ground state of moderately frustrated materials.

INTRODUCTION

Frustrated antiferromagnets (AFMs) have several degenerate—or nearly degenerate—ground states and generally do not order until very low temperature (i.e., low Néel temperature $T_N$) due to competing interactions caused by geometry, exchange interactions, or other factors. However, their Curie-Weiss temperature ($\Theta_{CW}$) is large and negative due to strong antiferromagnetic correlations. This juxtaposition led to the quantification of magnetic frustration as $f = |\Theta_{CW}/ T_N|$, where a large $f$ value indicates a highly frustrated system.[1] In the most extreme case where $T_N = 0$ and $f = \infty$, the material may be a quantum spin liquid (QSL). QSLs exhibit entanglement, a quantum mechanical correlation between electrons on different atoms. The long-range entanglement present in a QSL could aid in the development of quantum computers, and the spin liquid state could shield quantum qubits against outside noise.[2]

Slight changes in synthesis conditions can significantly influence the structure and spin states of materials, including quantum materials such as AFMs and QSLs. One reason for this may be because minor deviations in crystal structure affect the bond distances and exchange interactions between spins in magnetic structures. For example, small alterations in the synthesis recipe of the frustrated AFM barlowite $Cu_4(OH)_6FBr$, such as the precursor used, have been shown to affect its structure and magnetic properties.[3,4] It is therefore important to investigate more broadly how various synthesis conditions can influence the crystal structures and spin state(s) of quantum materials and how they interact with a host of other factors such as, composition, lattice, spin-orbit coupling (SOC) and J coupling, etc.

The most common synthetic variables include reagents, temperature, time, pH, and pressure. Magnetic field is an understudied synthetic parameter; most usage of a magnetic field has occurred as a post-synthetic treatment instead of during the initial synthesis. For example, application of a magnetic field during annealing is commonly performed to align ferromagnets and has also been known to change the properties of metals and alloys.[5,6] Nonetheless, magnetosynthesis, the use of a magnetic field during synthesis, is an emerging field of research and has been shown to impact the crystal structure and magnetic properties of a few iridates and ruthenates.[7]

Cao et al. found that magnetosynthesis using only a 0.02–0.06 T field affected the magnetic properties and crystal structures of the AFM insulators $Ba_4Ir_3O_{10}$, $Sr_2IrO_4$, and $Ca_2RuO_4$.[7] As-synthesized $Ba_4Ir_3O_{10}$ is a quantum liquid down to 0.2 K, yet magnetosynthesis induced long-range AFM order at the high temperature of $T_N$ = 125 K and relieved the magnetic frustration.[7,8] After magnetosynthesis at 0.02–0.06 T, the $T_N$ of $Sr_2IrO_4$ was reduced from 240 K to 150 K, and $Ca_2RuO_4$, which is normally an AFM with a $T_N$ of 110 K, became ferromagnetic.[7,9] Structural and electronic changes after magnetosynthesis were found as well, including reduced distortion in $Ca_2RuO_4$ and resistivity decreased by ~5 orders of magnitude in $Sr_2IrO_4$. Note that these compounds all exhibit strong SOC, which Cao et al. hypothesized is key to the effectiveness of magnetosynthesis.

Here, we investigate the effect of magnetosynthesis upon the composition, crystal structure, and magnetic properties of a series of $Cu^{2+}$-containing material exhibiting a range of magnetic frustration, from simple AFM (CuCl$_2$·2H$_2$O) to canted or frustrated/spin-glass AFMs ((Cu,Zn)$_3$Cl$_4$(OH)$_2$·2H$_2$O and atacamite Cu$_2$(OH)$_3$Cl, respectively) to a highly frustrated QSL material herbertsmithite (HBS, Cu$_3$Zn(OH)$_6$Cl$_2$). To our best knowledge, the investigation of magnetosynthesis effect on the systems with spin-only coupling is the first precedent in literature. We employ various low-temperature synthesis methods that permit the incorporation of a permanent magnet: hydrothermal synthesis, evaporation of a salt solution formed during hydrothermal synthesis, or dehydration-rehydration. CuCl$_2$·2H$_2$O is a simple AFM with a low ordering temperature $T_N$ = 4.3 K and a $\Theta_{CW}$ of around –5 K, so it is not frustrated.[10,11] The compound Cu$_3$Cl$_4$(OH)$_2$·2H$_2$O has only been reported twice;[12,13] it is AFM with a $T_N$ = 17.5 K and a $\Theta_{CW}$ of around +17.1, meaning this material is also not magnetically frustrated.[13] However, it exhibits a small net moment upon field cooling due to spin canting.[13] Atacamite (Cu$_2$(OH)$_3$Cl) is a frustrated AFM ($f \approx 17$) that exhibits spin glass behavior in synthetic samples.[14–19] HBS (Cu$_3$Zn(OH)$_6$Cl$_2$) is the leading QSL candidate due to its arrangement of Cu$^{2+}$ in an undistorted kagomé lattice.[20–22] We also explore the magnetosynthesis of mixed Cu/Zn versions for Cu$_3$Cl$_4$(OH)$_2$·2H$_2$O compound. As Zn$^{2+}$ is non-magnetic, Zn$^{2+}$ substitution onto Cu$^{2+}$ sites is expected to influence these materials' magnetic properties. This work explores the magnetic and structural properties of these compounds when synthesized with and without the presence of magnetic fields of strengths 0.09, 0.19 and 0.37 T.

EXPERIMENTAL SECTION

*Materials.* Cu$_2$(OH)$_2$CO$_3$ (54-56% Cu, Thermo Scientific), ZnCl$_2$ anhydrous (99.95%, Alfa Aesar), CuO (+99%, ThermoFisher), ZnO (99.0%, Alfa Aesar), CuCl$_2$·2H$_2$O (reagent grade, Sigma Aldrich), and HCl (ACS grade, Fisher Chemical).

*Synthesis of HBS $Cu_3Zn(OH)_6Cl_2$*. The reagents and stoichiometries shown in Table S1 were mixed in PTFE-liners and placed in hydrothermal autoclaves. The autoclaves were heated to 210 °C over 3 hours, held at 210 °C for 24 hours, and then cooled to room temperature over 30 hours. The product was filtered and washed with acetone.

*Evaporative crystallization of $(Cu,Zn)_3Cl_4(OH)_2\cdot 2H_2O$ and $CuCl_2\cdot 2H_2O$.* Three different trials were conducted, one with only Cu, one with only Zn, and one with a mixture of Cu and Zn. The reagents and stoichiometries shown in Table S2 were mixed in PTFE-liners and placed in hydrothermal autoclave. The autoclaves were heated to 210 °C over 3 hours, held at 210 °C for 24 hours, and then cooled to room temperature over 30 hours. Following the heating step, all solutions were placed in glass vials or petri dishes and left to evaporate over the course of weeks.

*Synthesis of acatamite $Cu_2(OH)_3Cl$.* The $CuCl_2\cdot 2H_2O$ reagent was heated to 200 °C, held at 200 °C for 72 hours, and then cooled to room temperature. The resulting yellow powder was left in a vial to hydrate over the course of weeks. The final product was washed with acetone and filtered.

*Single-crystal X-ray diffraction (SCXRD).* We determined the crystal structures of the products with SCXRD using a Bruker D8 Venture with a Ga MetalJet source (1.341 Å) at 130 K and room temperature. The structures were solved with intrinsic phasing in APEX6 and refined with SHELXL and OLEX2.[23–27] Hydrogen atoms were inserted at positions of electron density near the oxygen atoms and were refined with a fixed bond length and an isotropic thermal parameter 1.5 times that of the attached oxygen atom. Thermal parameters for all other atoms were refined anisotropically.

*Powder X-ray diffraction (PXRD).* For PXRD, products were ground using a mortar and pestle and placed on silicon zero-background slides with grease. PXRD measurements were taken using a Rigaku Ultima IV and Rigaku SmartLab in Bragg-Brentano geometry with Cu $K_\alpha$ radiation. For Rigaku Ultima, we used a 10 mm slit, a K$_\beta$ filter, and a 0.02° step with a 0.5 hour collection (or 8-12 hour for scans used in the Rietveld refinement). For Rigaku SmartLab, we used a 10 mm slit, a K$_\beta$ filter, and a 0.001° step size with a 3.5 hour collection. Rietveld refinements were performed using TOPAS.[28] Hydrogen atoms were excluded from the refinement model.

*Magnetism.* AC susceptibility and DC magnetization measurements were performed using a Quantum Design Physical Properties Measurement System (PPMS) Dynacool. AC susceptibility was measured with a 1.5 Oe drive current at a frequency of 10,000 Hz. DC magnetization measurements were performed under an applied field of 0.1 T in the temperature range of 2−350 K. For the FC measurements, the sample was cooled down under an applied field of 0.1 T. Magnetization measurements were performed in an applied field ranging from −14 to 14 T.

*Scanning electron microscopy (SEM) with the Energy-dispersive spectroscopy (EDS) adapter.* SEM imaging was done on a Hitachi S-4800 SEM operating at a 15 keV accelerating voltage and 10 μA beam current. Elemental analysis was conducted by EDS on the same instrument using the included Pathfinder analysis software for quantification. Spectra were acquired for 60 s. EDS mapping was performed by acquiring data for 6 minutes.

*First-Principles Calculations.* Spin-polarized (collinear) Density functional theory (DFT) calculations were conducted to compute the quantum mechanical energies for the 2×2×2 supercell of $(Cu,Zn)_3Cl_4(OH)_2 \cdot 2H_2O$ using the Vienna Ab-Initio Package (VASP),[29,30] with the Projector Augmented-Wave method[31] using the Perdew-Burke-Ernzerhof (PBE) generalized gradient approximation.[32] DFT basis cut-off energies were 520 eV and the Gamma point was used for the k-point mesh. $(Cu,Zn)_3Cl_4(OH)_2 \cdot 2H_2O$ exhibits canted AFM, but in the absence of experimental information on its magnetic propagation vector, all calculations were performed assuming a collinear ferromagnetic ordering. The formation energies were referenced to the elemental DFT free energies provided by the Materials Project[33] and adjusted using the MaterialsProject2020Compatibility scheme[34] implemented in Pymatgen.[35]

RESULTS

### A. Magnetosynthesis of HBS

As HBS $Cu_3Zn(OH)_6Cl_2$ is the leading experimental kagome QSL candidate material,[20–22] we initially focused our hydrothermal synthesis and magnetosynthesis attempts on this phase. However, it can be challenging to synthesize pure HBS due to the tendency of Cu to substitute on the interlayer Zn sites, and even the purest phases tend to have ~15% Cu on the Zn sites (i.e., $Cu_3Zn_{0.85}Cu_{0.15}(OH)_6Cl_2$).[36,37] As a starting point, we used $Cu_2(OH)_2CO_3$ and anhydrous $ZnCl_2$ (see Table S1 for details) mixed with water and heated in a PTFE-lined autoclave at 210 °C for 24 hours. A large excess of $ZnCl_2$ was required to eliminate tenorite (CuO) impurities identified via PXRD (Figure S1), consistent with prior attempts.[38]

The optimized synthesis was repeated both with and without a SmCo permanent magnet under the autoclave, which yielded a magnetic field of 0.09 T at the bottom of the PTFE liner. Throughout the manuscript, we label samples with the field used to synthesize them as, e.g., HBS_0T and HBS_0.09T. Both products were pure according to PXRD and visual inspection (Figure 1a). We performed LeBail fits of the PXRD data for each sample, and the extracted lattice parameters were identical within error (Figure S8).

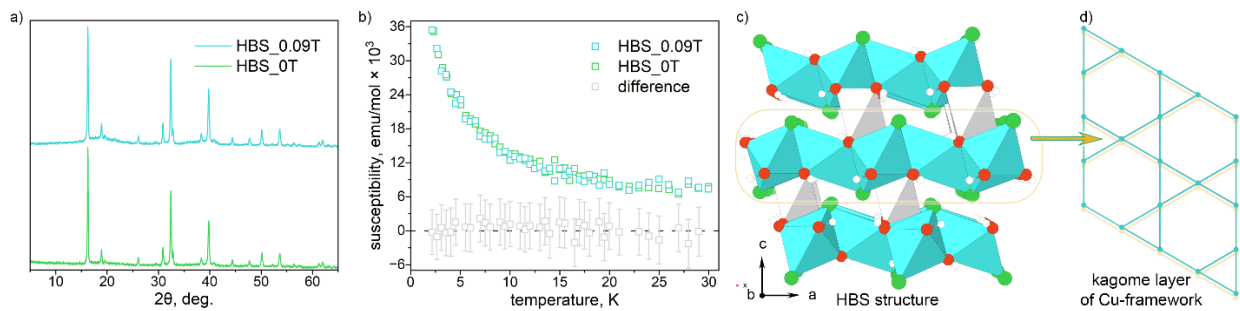

Figure 1. (*a*) PXRD data comparing the HBS synthesis under 0.09 T (blue) and 0 T (green). (*b*) ZFC magnetic susceptibility of HBS synthesized with (blue) and without 0.09 T magnetic field (green) as well as the difference between those two (gray). (*c*) View of HBS structure and (*d*) kagome layer built from the Cu-framework. White, red, green, blue, and gray spheres, blue and gray polyhedra represent H, O, Cl, Cu, and Zn atoms, Cu and Zn octahedra, respectively.

The low temperature magnetic susceptibility data of HBS synthesized under a 0.09 T magnetic field were indistinguishable from that of HBS synthesized with no field (Figure 1b). The data for both samples are consistent with QSL behavior: there is no magnetic transition down to 2 K. For both the 0.09 T and the 0 T syntheses, the zero field-cooled (ZFC) and field-cooled (FC) susceptibilities were identical (Figures S18, S19), matching the results found by Shores et al. for HBS synthesized without a magnetic field.[20] Moreover, Curie-Weiss fits in the 100–350 K temperature range resulted in similar Curie-Weiss temperature ($\Theta_{CW}$) and effective moment ($\mu_{eff}$) for $Cu^{2+}$ (Table S15 and Figure S23) corroborating that at least 0.09 T field is not enough to have a strong effect in HBS magnetosynthesis.

### B. Crystal Growth of Metastable Cu-Containing Chlorides

Next, we explored synthesis of HBS using CuO and $ZnCl_2$ reagents hoping to grow single crystals and potentially increase the yield due to impact on the CuO vs. $Cu_2(OH)_2CO_3$ dissolution kinetics in hydrothermal conditions. However, it was more difficult to produce a pure product using CuO than $Cu_2(OH)_2CO_3$: at a 0.08:1 Cu:Zn ratio the product was still contaminated with CuO impurities, and only extreme ratios of 0.04:1 produced small amounts of pure product at 5% yield (Table S1). Therefore, we added concentrated HCl to the hydrothermal reactions. At a high enough concentration, HCl successfully dissolved CuO; however, the HBS powder was also dissolved, and no product precipitated. We then attempted to use these clear solutions in evaporative crystallization under no and small magnetic fields at room temperature to grow single crystals of Zn/Cu/Cl-containing materials. To compare the effect(s) of $Cu^{2+}$ and $Zn^{2+}$ ions on crystallization dynamics, we prepared aqueous solutions of Cu and/or Zn chlorides with HCl (see Table S2 for details) and heated them in a PTFE autoclave at 210 °C for 24 hours. The solutions were then permitted to evaporate at room temperature in petri dishes.

Immediately after the heating step, the Zn-only solution was clear. Upon evaporation for 3-5 weeks, a clear liquid layer remained; no crystals formed (Figure 2). However, the Cu-only solution was dark green after the heating step. After a few days, the liquid evaporated fully and loose clusters of light blue flaky crystals remained. Unfortunately, these crystals were not suitable for SCXRD; therefore, this phase was identified as $CuCl_2·2H_2O$ by PXRD.

The most interesting case was presented by the mixed Cu/Zn solution: after heating, it was a bright/light blue color. After a few days, the solution became teal green/blue (Figure 2). Once the liquid was almost evaporated, blue crystals formed. Several weeks later, as the liquid continued to evaporate, blue and green crystals were both present. Interestingly, we could see the intergrowth between green and blue crystals in an optical microscope. Yet a couple weeks after that, only green crystals remained. We note that the liquid never fully evaporated; a mother liquor was retained surrounding the crystals. The blue crystals were long needles, sometimes even reaching a few millimeters in length. The green crystals were smaller and exhibited an irregular hexagonal plate-like morphology. Through SCXRD, the blue crystals were identified as $(Cu,Zn)Cl_2·2H_2O$ (isostructural to $CuCl_2·2H_2O$ as unit cell check suggested), and the green crystal structure was solved as $(Cu,Zn)_3Cl_4(OH)_2\cdot 2H_2O$ (isostructural to $Cu_3Cl_4(OH)_2\cdot 2H_2O$ in space group $P\text{–}1$, Tables S4–S8); this structure is described in more detail below.

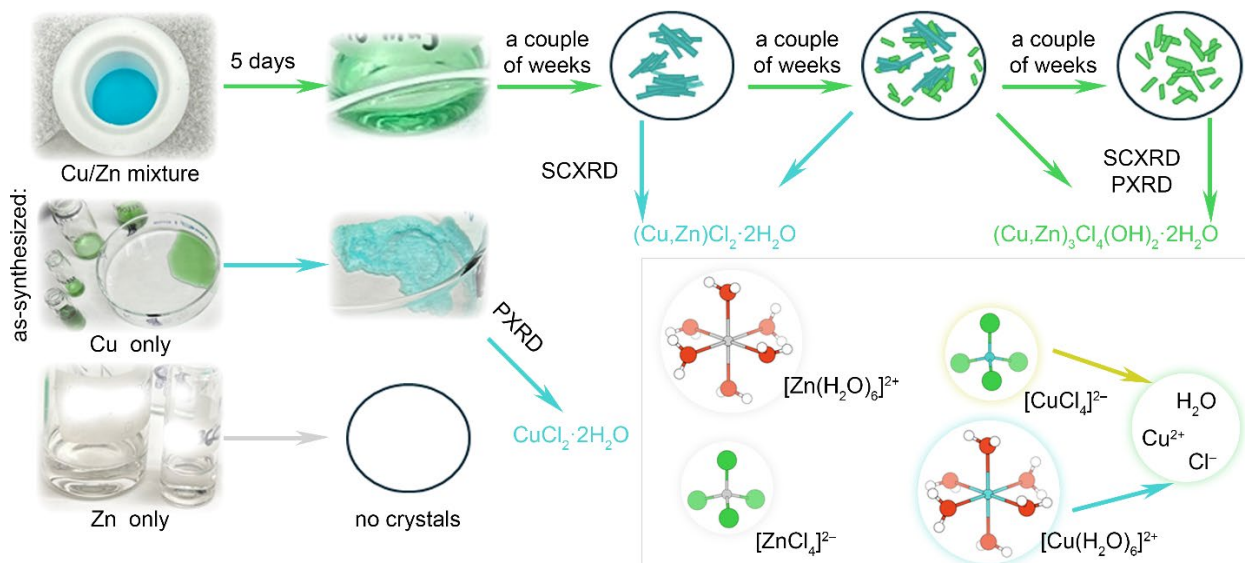

Figure 2. Evaporative crystallization from Cu and/or Zn solutions after hydrothermal reaction. The bottom right insert shows the main complexes present in the solution after hydrothermal reaction. White, red, green, blue, and gray spheres, H, O, Cl, Cu, and Zn atoms, respectively. The color-coding of the complexes corresponds to their appearance in solution.

Despite the similarity of $Cu^{2+}$ and $Zn^{2+}$ in terms of oxidation state and ionic radius (0.73 Å and 0.74 Å for $Cu^{2+}$ and $Zn^{2+}$ in octahedral coordination, respectively[39]), the results of these evaporations varied widely. One major difference between $Cu^{2+}$ ($d^9$) and $Zn^{2+}$ ($d^{10}$) is that $Cu^{2+}$ experiences Jahn-Teller effects due to its unpaired electron, whereas $Zn^{2+}$ does not. We speculate that these effects lead to the instability or metastability of the blue $(Cu,Zn)Cl_2\cdot 2H_2O$ phase — it slowly transforms to the green $(Cu,Zn)_3Cl_4(OH)_2\cdot 2H_2O$ phase, which thus is more thermodynamically stable. SEM images of the blue and green crystals show that the blue crystal is more granular than the green, supporting the hypothesis of $(Cu,Zn)Cl_2\cdot 2H_2O$ instability (Figure S3). We hypothesize this is likely because the lattice does not easily accommodate $Zn^{2+}$ in the irregular octahedral sites. Additionally, upon tallying known compounds involving zinc and copper coordination, $Zn^{2+}$ coordinated to $Cl^-$ tends to prefer a tetrahedral geometry, whereas $Cu^{2+}$ coordinated to Cl⁻ prefers other geometries: tetrahedral, square planar, square pyramid, trigonal pyramid and octahedral (Table S3). In $CuCl_2·2H_2O$, $Cu^{2+}$ is coordinated in an octahedral geometry to four Cl⁻ and two $H_2O$, which may contribute to the lack of stability of $Zn^{2+}$ in these sites (see Figure 3a). Apart from that, the differences and progression in the solution color tells the information about $Cu^{2+}$ coordination environment in aqueous media. For the only $Cu^{2+}$ solution, after hydrothermal dissolution of CuO in HCl the solution is green, indicating the presence of $[CuCl_4]^{2-}$ tetrahedral complexes because a mixture of blue $[Cu(H_2O)]_6^{2+}$ and yellow $[CuCl_4]^{2-}$ results in green color (Figure 2).[40] After evaporative crystallization, this solution turned blue, corroborating the evaporation of residual HCl and Cl⁻ being incorporated into $CuCl_2·2H_2O$ structure, hence leaving the blue colored $[Cu(H_2O)]_6^{2+}$ ions in solution (Figure 2).[40] Opposite to that, hydrothermal dissolution of CuO and $ZnCl_2$ leads to $[Cu(H_2O)]_6^{2+}$ species in the aqueous media resulting in blue color potentially due to preferential $Zn^{2+}$ bonding to Cl⁻. This solution later turns green; therefore, it is likely to contain $[CuCl_4]^{2-}$ and not just octahedral $Cu^{2+}$, since the presence of $Zn^{2+}$ might preserve more Cl⁻ ions in solution and let $H_2O$ evaporate slower, producing better quality crystals (Figures 2 and S4–S6).

C. *Crystal Structures of $(Cu,Zn)_3Cl_4(OH)_2·2H_2O$ Obtained under No and Weak Magnetic Field*

The $Cu_3Cl_4(OH)_2·2H_2O$ structure has only two references in the literature.[12,13] One of the previous synthetic approaches involved the solid state reaction between $CuCl_2·2H_2O$ and $Cu(OH)_2$ powders in air at 75 °C for 5 days followed by washing with alcohol and drying at 75°C.[12] Another route consisted of a non-trivial cycle: dehydration of $CuCl_2·2H_2O$ at 200 °C for several weeks, rehydration at room temperature, and wetting resulting powder with a couple of water drops followed by drying at 50–60 °C.[13] In contrast to our $(Cu,Zn)_3Cl_4(OH)_2·2H_2O$ crystal growth method, only powders of $Cu_3Cl_4(OH)_2 \cdot 2H_2O$ were produced by Walter-Lévy et al. in 1970 and Asaf et al. in 1996.[12,13] The authors also mentioned that it was challenging to obtain crystalline and phase-pure material.[13]

Here, we took advantage of the room-temperature evaporative crystal growth of $(Cu,Zn)_3Cl_4(OH)_2 \cdot 2H_2O$ to probe the effect of a 0.19 T magnetic field upon the resulting material. In both synthetic routes, we obtained green block-shaped crystals after a couple of weeks and employed SCXRD to determine that the crystals were $(Cu,Zn)_3Cl_4(OH)_2 \cdot 2H_2O$ (sp. gr. $P$–1). This phase is isostructural to $Cu_3Cl_4(OH)_2 \cdot 2H_2O$, which has only been characterized by PXRD.[12,13] We therefore present here the first full structure determination of this type. $(Cu,Zn)_3Cl_4(OH)_2 \cdot 2H_2O$ is a layered structure accommodating four different $Cu^{2+}/Zn^{2+}$ sites coordinated by $Cl^-$, $OH^-$, and $H_2O$ ligands. As shown in Figure 3a, Cu1, Cu3, and Cu4 are in distorted octahedral geometries, while Cu2 is in distorted square pyramidal geometry. The layers in $(Cu,Zn)_3Cl_4(OH)_2 \cdot 2H_2O$ (Figures 3b and 3c) are built by chains of Cl,Cl-edge sharing $[CuCl_4(OH)_2]$ octahedra with the Cu3 and Cu4 alternating within the 1D moiety. Those chains are connected by Cl,Cl-edge sharing Cu2 dimers, i.e., $[CuCl_3(OH)_2]_2$, via Cl,OH-edge sharing to both Cu3 and Cu4 octahedra to create a rectangular mesh (Figure 3b). And finally, this sheet is decorated by the terminal Cu1 octahedra, $[CuCl_2(OH)_2(H_2O)_2]$, through OH,OH-edge sharing with the Cu2-containing square pyramid and Cl,OH-edge sharing with Cu3 and Cu4 octahedra (Figure 3b).

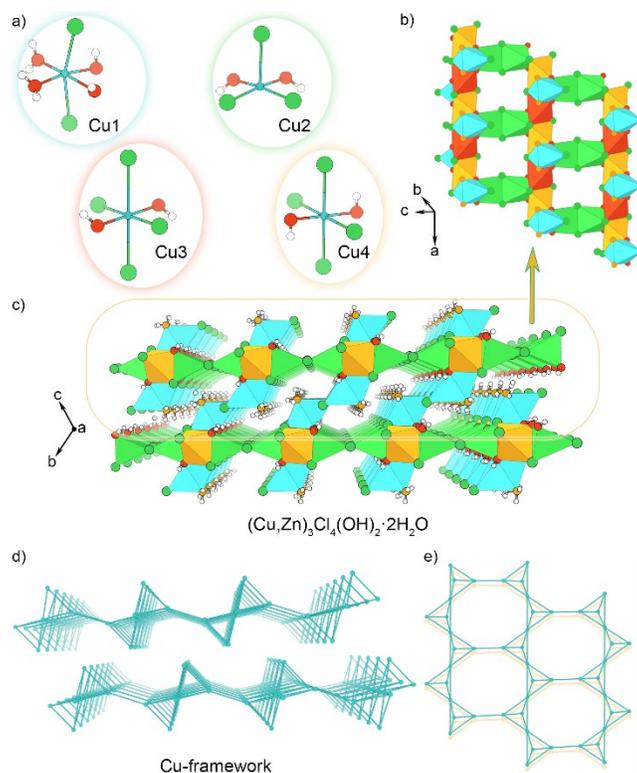

Figure 3. View of (*a*) the building polyhedral units, (*b*) the $(Cu,Zn)_3Cl_4(OH)_2 \cdot 2H_2O$ layers, (*c*) the overall $(Cu,Zn)_3Cl_4(OH)_2 \cdot 2H_2O$ structure, and (*d*, *e*) Cu framework. White, red, green, blue, and orange spheres represent H, O, Cl and Cu atoms and water molecules, respectively. Blue, green, red, and orange polyhedral represent Cu1, Cu2, Cu3, and Cu4, respectively. Zn atoms in the structure are omitted for clarity.

Even though the $(Cu,Zn)_3Cl_4(OH)_2 \cdot 2H_2O$ layers are charge-neutral, there is H···O and H···Cl hydrogen bonding leading to the layer stacking in the $(Cu,Zn)_3Cl_4(OH)_2 \cdot 2H_2O$ structure (Figure S10). Moreover, the terminal Cl1 ligand coordinated to Cu1 is significantly further from Cu2 in the adjacent layer (3.229 Å at 300 K), having a weak ionic interaction further gluing the layers together (Figure S11). However, this layered structure did not demonstrate signs of exfoliation as shown in the SEM image (Figures S4–S6), and it does not easily dissolve in water, isopropanol, or acetone primarily due to the strong hydrogen bonding and weak ionic connection between layers.

We note that SCXRD cannot conclusively distinguish $Cu^{2+}$ and $Zn^{2+}$ due to their nearly identical *Z* and ionic radii. However, we attempted to refine different occupancy models against the SCXRD data, which resulted in empirical formulae of $Cu_{2.86}Zn_{0.14}Cl_4(OH)_2 \cdot 2H_2O$ (RT, 0 T), $Cu_{2.87}Zn_{0.13}Cl_4(OH)_2 \cdot 2H_2O$ (130 K, 0 T), and $Cu_{2.93}Zn_{0.07}Cl_4(OH)_2 \cdot 2H_2O$ (130 K, 0.019 T), although the R-values were not significantly improved. The EDS results for $(Cu,Zn)_3Cl_4(OH)_2 \cdot 2H_2O$ crystal synthesized without an applied field suggested a Cu:Zn:Cl ratio of approximately 2.85:0.15:5.01 (Figure S7), and we thus used this Cu:Zn ratio to finalize the $(Cu,Zn)_3Cl_4(OH)_2 \cdot 2H_2O$ crystal structures.

Interestingly, the only location for $Zn^{2+}$ to substitute that did not result in refinement instability or negative occupancies was on the Cu1 site. Our hypothesis is that this position is more stable because only two $Cl^-$ are bonded to Cu1 in this octahedral coordination environment, as opposed to four, and $Cl^-$ ligands with $Zn^{2+}$ would usually prefer to be tetrahedral but substitution of $Cl^-$ to $H_2O$ or $OH^-$ shifts $Zn^{2+}$ geometry to octahedral coordination (Table S3). We therefore posit that $Zn^{2+}$ substitution destabilizes $CuCl_2 \cdot 2H_2O$, in which the metal site has octahedral coordination, and promotes formation of the $Cu_3Cl_4(OH)_2 \cdot 2H_2O$ structure with diverse octahedral metal sites. These hypotheses explain why the $(Cu,Zn)_3Cl_4(OH)_2 \cdot 2H_2O$ crystal is more thermodynamically stable in the Cu/Zn mixture trial.

To probe our hypothesis that Zn substitution stabilizes the $(Cu,Zn)_3Cl_4(OH)_2 \cdot 2H_2O$ structure, we performed spin-polarized DFT calculations on the 2×2×2 supercell (see Experimental Section for more details). The parent $Cu_3Cl_4(OH)_2 \cdot 2H_2O$ structure contains four crystallographically distinct Cu sites. To assess the site preference and energetic impact of Zn substitution, we generated four $Cu_{2.5}Zn_{0.5}Cl_4(OH)_2 \cdot 2H_2O$ structures in which Zn replaced each Cu site respectively (e.g Zn1=Zn@Cu1). Comparing the resulting formation energies from the relaxed DFT calculations reveals that among the substituted structures, Zn substitution on the Cu1 site yields the lowest formation energy, computed relative to the elemental DFT free energies from the Materials Project,[33,34] consistent with the SCXRD results. To further investigate Zn stabilization in (Cu,Zn)$_3$Cl$_4$(OH)$_2$·2H$_2$O, we enumerated the 2×2×2 supercell with partial occupancies on the Cu1 site and performed spin-polarized DFT relaxations on the symmetrically-unique structures. The convex hull (Figure 4b), which is the stability line connecting the most thermodynamically favored Cu$_{3-x}$Zn$_x$Cl$_4$(OH)$_2$·2H$_2$O compositions, shows that Zn substitution at the Cu1 site progressively stabilizes the structure as the Zn fraction (*x*) increases. Compositions that lie on the convex hull are predicted to be thermodynamically stable at 0 K, while those slightly above the hull are metastable but may still be experimentally accessible under magnetic-field-induced stabilization, and 'quenched' in a remanently metastable state after the field is removed.[41] The lowest Zn fraction (x) for the Cu$_{3-x}$Zn$_x$Cl$_4$(OH)$_2$·2H$_2$O structures lying on the convex hull is x = 0.3, approximately twice the Zn concentration determined by EDS (x = 0.15). This discrepancy likely arises from the uncertainties in the EDS analysis and the computational constraints inherent to DFT, which restrict the compositional resolution that can be modeled and thus may shift the predicted onset of stability. Overall, the DFT calculations corroborate the preferential occupation of Zn at the Cu1 site and provide insight into the thermodynamic stability of Cu$_{3-x}$Zn$_x$Cl$_4$(OH)$_2$·2H$_2$O.

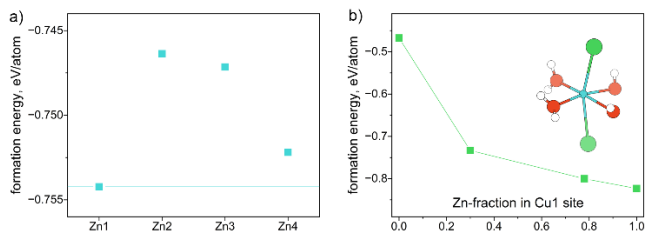

**Figure 4.** (*a*) Formation energies of the Cu$_{2.5}$Zn$_{0.5}$Cl$_4$(OH)$_2$·2H$_2$O structures with Zn only on Cu1, Cu2, Cu3, and Cu4 sites. (*b*) Formation energies of Cu$_{3-x}$Zn$_x$Cl$_4$(OH)$_2$·2H$_2$O structures where x is Zn fraction or Zn occupancy on the Cu1 site. Note: only Cu$_{3-x}$Zn$_x$Cl$_4$(OH)$_2$·2H$_2$O compositions that are on the convex hull, are shown on the graph. The insert shows Cu1 site coordination environment. White, red, green, and blue spheres represent H, O, Cl and Cu atoms respectively.

Magnetosynthesis of (Cu,Zn)$_3$Cl$_4$(OH)$_2$·2H$_2$O crystals with a 0.19 T applied field resulted in the same product as without a magnetic field with nearly identical unit cells (Table S4); the lattice parameters differ by approximately 0.2–1%. We speculated that magnetic field would affect the most magnetic ions like $Cu^{2+}$. Comparison of Cu coordination environment between crystals synthesized with and without applied magnetic field revealed that the most affected site is Cu1 (Figure 5a and Tables S5–S8). The changes in the bond distances and angles in the Cu1 octahedron span within 0–0.5%, with bigger changes outside of 3σ range (Tables S7–S8). This change in Cu1 coordination propagates to a Cu1 shift within the Cu network. However, as Cu1 is also the site that we hypothesize can accommodate $Zn^{2+}$ most easily (as discussed above), we cannot rule out the possibility that these changes might also be due to different Zn incorporation levels in different crystals.

The Cu atoms themselves form an interesting lattice, shown below in Figures 3d and 3e. This lattice has stretched octagons in roughly the (011) plane formed by six tetrahedra (Figure 3d). These tetrahedra are highly distorted, as shown in Figure 5b. From the side, these distorted octagons stack as layers of interconnected bowties formed by Cu tetrahedra (Figure 5b). The Cu1–Cu2–Cu4 and Cu1–Cu3–Cu4 triangles are affected the most by magnetic field, and they shrink by 0.1-0.2%; this change is in 2σ range (Table S7).

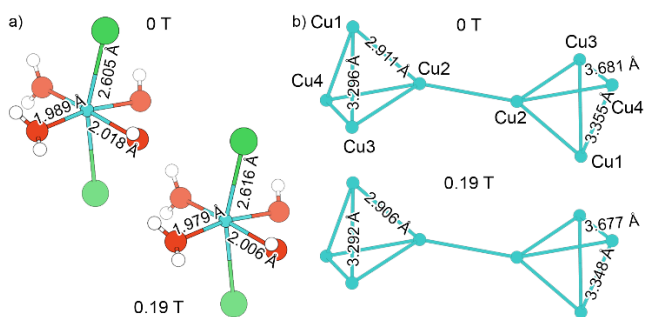

**Figure 5.** View of (*a*) Cu1 octahedra and (*b*) Cu network. Interatomic distances are shown for the most significant change (>0.10%) between the $(Cu,Zn)_3Cl_4(OH)_2 \cdot 2H_2O$ crystal structures solved at 130 K and grown with and without a 0.19 T field. White, red, green, and blue spheres represent H, O, Cl and Cu atoms, respectively.

*D. Magnetic Properties of the $(Cu,Zn)_3Cl_4(OH)_2 \cdot 2H_2O$ Phase*

We performed a Rietveld refinement on PXRD data of several ground $(Cu,Zn)_3Cl_4(OH)_2 \cdot 2H_2O$ green crystals synthesized without an applied magnetic field (Figures 6a and S9; Tables S9–S10). The structural model fit the data well ($R_w$ = 3.40%), and no impurity phases were observed. We note that one prior report of polycrystalline $Cu_3Cl_4(OH)_2 \cdot 2H_2O$ contained $CuCl_2 \cdot 2H_2O$ impurities, which are not present in our sample.[13]

We postulate that the tetrahedra of magnetic $Cu^{2+}$ cations in $(Cu,Zn)_3Cl_4(OH)_2 \cdot 2H_2O$ may be magnetically frustrated, as is common in materials with a kagomé or pyrochlore lattice, and that the magnetic behavior is highly correlated to the type and level of structural distortion in the lattice. Therefore, we investigated the magnetic properties of this compound by measuring the magnetic susceptibility of a collection of small crystals of $(Cu,Zn)_3Cl_4(OH)_2 \cdot 2H_2O$ totaling ~1 mg. We note that it was difficult to obtain a large enough quantity of these crystals for high quality measurements; in fact, we could not harvest even 1 mg of $(Cu,Zn)_3Cl_4(OH)_2 \cdot 2H_2O$ crystallized under a magnetic field and therefore could not perform PXRD or magnetic characterization.

As Asaf et al. found for $Cu_3Cl_4(OH)_2 \cdot 2H_2O$,[13] the susceptibility data of $(Cu,Zn)_3Cl_4(OH)_2 \cdot 2H_2O$ display antiferromagnetic ordering at low temperature, with some hysteresis and a small net moment likely due to canting of the spins (Figure 6b). We observe a broad peak in the temperature-dependent susceptibility data collected at low field ($\mu_0H$ = 0.005 T, Figure 6a) with splitting between the ZFC and FC data, which may indicate some spin glass character. Based on both DC magnetization and AC susceptibility measurements (Figures 6b and 6c, respectively), the Néel temperature ($T_N$) of $(Cu,Zn)_3Cl_4(OH)_2 \cdot 2H_2O$ is approximately 15.5 K, slightly lower than the value of 17.5 K reported for $Cu_3Cl_4(OH)_2 \cdot 2H_2O$;[13] this is reasonable given the substitution of non-magnetic $Zn^{2+}$ for magnetic $Cu^{2+}$. The field-dependent data show a small hysteresis loop with a small net moment that is likely due to canting of the spins (Figure 6d). Due to the small amount of sample, there was a large diamagnetic component in the data from the sample holder (Figure S20), and a linear Curie-Weiss fit was not possible even with a diamagnetic correction ($\chi_0$).

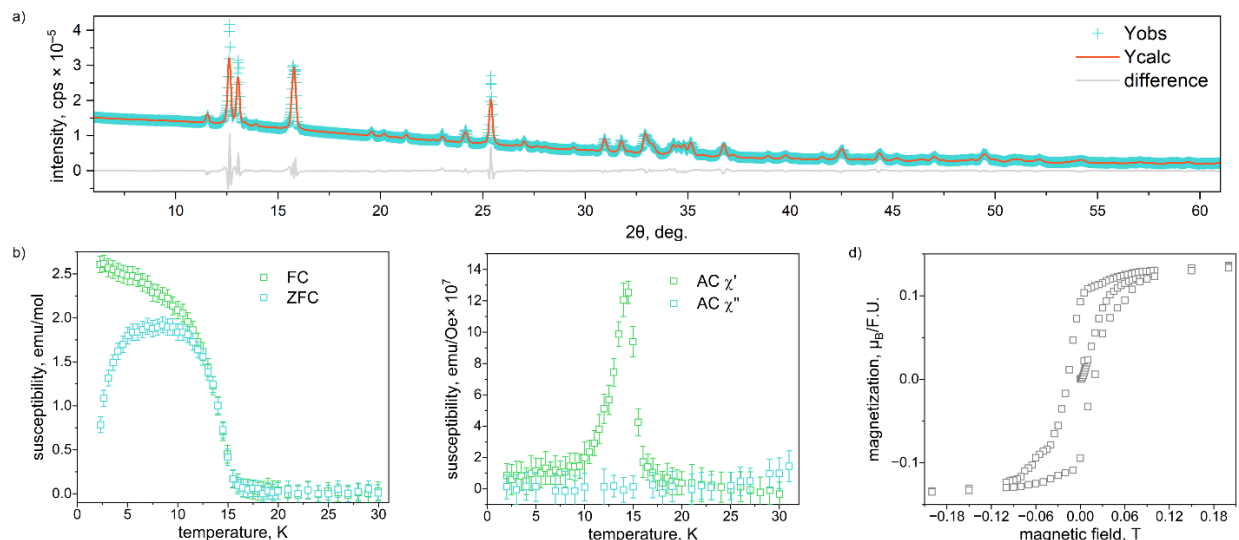

**Figure 6.** (*a*) Rietveld refinement of a few (Cu,Zn)$_3$Cl$_4$(OH)$_2$·2H$_2$O crystals. Note: the difference (*b*) DC magnetization and (*c*) AC magnetic susceptibility (2−30 K), and (*d*) MvH plot at 2 K for a few (Cu,Zn)$_3$Cl$_4$(OH)$_2$·2H$_2$O crystals. ZFC and FC data are shown in blue and green, respectively. Real and imaginary components of AC susceptibility shown in blue and green, respectively. Note: the error bars in the MvH plot are smaller than the square size.

*E. Magnetosynthesis of CuCl$_2$·2H$_2$O*

Since we were not able to investigate the effect of magnetosynthesis on the magnetic properties of the frustrated (Cu,Zn)$_3$Cl$_4$(OH)$_2$·2H$_2$O structure due to low synthetic yield, our next effort to probe the effect of magnetosynthesis on d$^9$ (*S*=1/2) electronic systems focused on the room temperature evaporative crystallization of CuCl$_2$·2H$_2$O.

The solution obtained via hydrothermal CuO/CuCl$_2$ dissolution in HCl was allowed to evaporate under no magnetic field and fields of 0.19 T and 0.37 T to obtain blue powder of CuCl$_2$·2H$_2$O. The phase purity of all CuCl$_2$·2H$_2$O samples was confirmed by PXRD. For CuCl$_2$·2H$_2$O at 0 T, we performed a Rietveld refinement of the data collected at room temperature and found good agreement with the reported structure in *Pmna* space group ($R_w$ = 2.22%, 2.59%, and 2.60% for 0 T, 0.19 T, and 0.37 T, respectively, Figures S13–S15; Tables S11–S12); no additional phases were indexed. The changes in the unit cell for compositions synthesized under different applied field vary within 1σ range which is insignificant.

The $CuCl_2 \cdot 2H_2O$ structure consists of edge-sharing $[CuCl_4(H_2O)_2]$ octahedral chains with terminal water molecules (Figures 7a and 7b). The Cu network is represented by linear chains (Figure 7c) aligned along (0 0 1), indicating no prerequisite for frustration based on triangular geometry. The DC magnetization data of all three samples (0 T, 0.9 T, and 0.37 T) revealed a broad AFM peak at around 5.8 K (Figure 7d, Table S16) with no significant discrepancy between ZFC and FC data (Figures 7d and 7e). This matches the literature which lists the AFM transition at 4.3 K (inflection point).[10] The real part of the AC susceptibility (χ', Figure 7f) for all three samples demonstrated a broad peak centered at 5.7 K, matching the DC magnetization data.

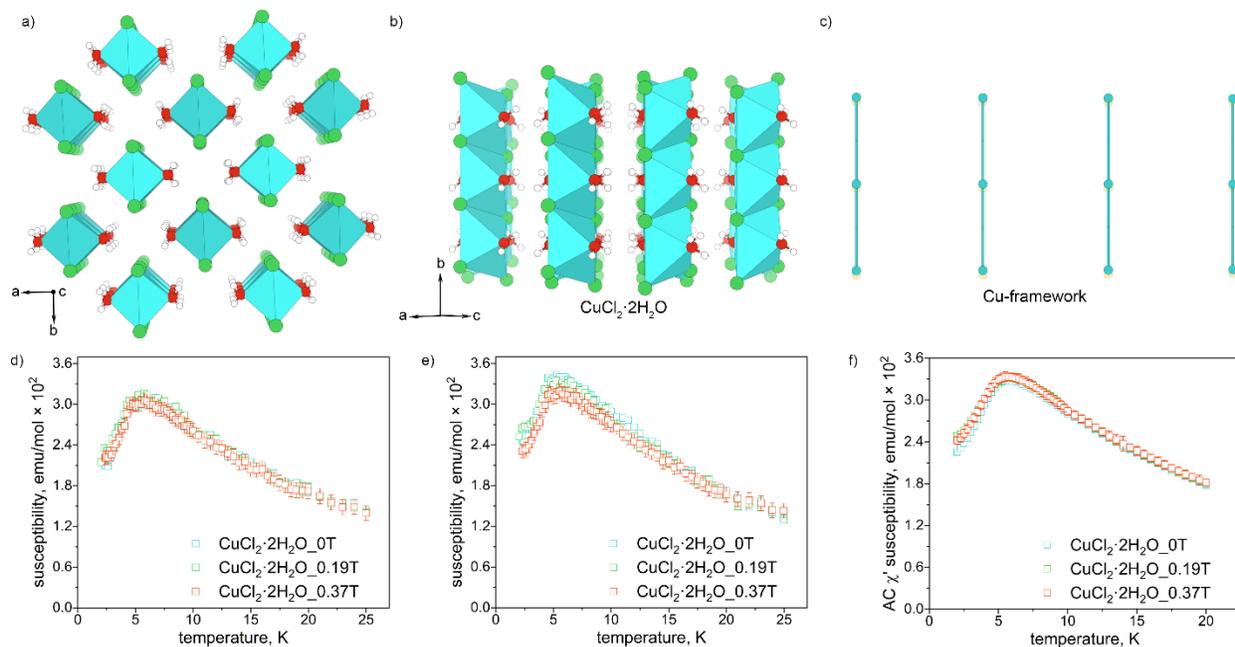

**Figure 7.** View of the $CuCl_2 \cdot 2H_2O$ structure along (*a*) [001] and (*b*) [101] and (*c*) Cu framework. White, red, green, and blue spheres and blue polyhedral represent H, O, Cl and Cu atoms, and Cu octahedra, respectively. (*d*) ZFC and (*e*) FC DC magnetization, and (*f*) AC magnetic susceptibility (2−25 K) for the $CuCl_2 \cdot 2H_2O$ powder synthesized under various magnetic field strengths: 0 T (blue), 0.19 T (green), and 0.37 T (red), respectively. The error bars are shown only for the $CuCl_2 \cdot 2H_2O\_0.37T$ sample for clarity. Only real components are shown for AC susceptibility.

As shown in Figures 7d–7f, there is no apparent difference in the magnetic susceptibility of $CuCl_2·2H_2O$ synthesized under varying magnetic field strengths. Each curve was smoothed with a weighted adjacent-averaging algorithm, and the peak maximum was determined as the 1$^{st}$ derivative plot crossing $y$=0. These results are shown in Table S16 and indicate that there is not a significant difference in AFM peak between samples. The high temperature magnetic susceptibility data were fit to a Curie-Weiss model. As shown in Table S16, synthesis under a magnetic field influenced $\Theta_{CW}$; the values increase (i.e., become less AFM) with the field applied during the synthesis: –10.8(8) K, –3.0(10) K, and 2.3(7) K for 0 T, 0.19 T, and 0.37 T, respectively. Due to the broad nature of the AFM transition in $CuCl_2·2H_2O$, it is hard to quantitatively compare Néel temperature or inflection point between three samples. Unfortunately, the discrepancy in the Weiss temperature is not strong enough evidence on its own to conclude whether there is a magnetosynthesis effect on $CuCl_2·2H_2O$.

### F. Magnetosynthesis of Atacamite $Cu_2(OH)_3Cl$

Finally, our attempts to synthesize $Cu_3Cl_4(OH)_2·2H_2O$ via the published $CuCl_2·2H_2O$ rehydration methods yielded instead a mixture of $Cu_3Cl_4(OH)_2·2H_2O$ and atacamite $Cu_2(OH)_3Cl$.[13] Over the course of several weeks, $Cu_3Cl_4(OH)_2·2H_2O$ transformed to $Cu_2(OH)_3Cl$ (Figure S2), confirming the metastability of the all-Cu $Cu_3Cl_4(OH)_2·2H_2O$ structure.

Atacamite $Cu_2(OH)_3Cl$ has an orthorhombic structure with a distorted triangular lattice of $Cu^{2+}$ forming a weakly coupled 3D network of anisotropic sawtooth chains,[14] which leads to geometric magnetic frustration (Figures 8a and 8b). The low temperature magnetic behavior of both natural and synthetic atacamite have been the subject of prolonged interest and debate: it exhibits a magnetic transition to an AFM but likely disordered and/or spin-glassy ground state.[15,16,18,19,42,43] While different ordering temperatures of $T_N \approx 5.3$ K for synthetic atacamite[15,16] and $T_N \approx 9$ K for natural atacamite[18,19] have been reported, both materials are highly frustrated $S=1/2$ quantum magnets ($f \approx 17$ for synthetic samples[16]).

We synthesized polycrystalline atacamite in both no magnetic field and an applied field of 0.19 T (. While these samples were not highly crystalline due to the synthetic process, we performed Rietveld fits of PXRD data collected at room temperature (Figures S16–S17). The sample synthesized without a magnetic field yielded pure $Cu_2(OH)_3Cl$ atacamite product ($R_w$ = 3.75%), while the $Cu_2(OH)_3Cl\_0.19T$ sample had a small impurity of $Cu_3Cl_4(OH)_2·2H_2O$ at 0.3 wt.% ($R_w$ = 3.87%). The unit cell and bond distance differences between models for the $Cu_2(OH)_3Cl\_0T$ and $Cu_2(OH)_3Cl\_0.19T$ samples are within the 1σ range which is linked to limitations of PXRD analysis.

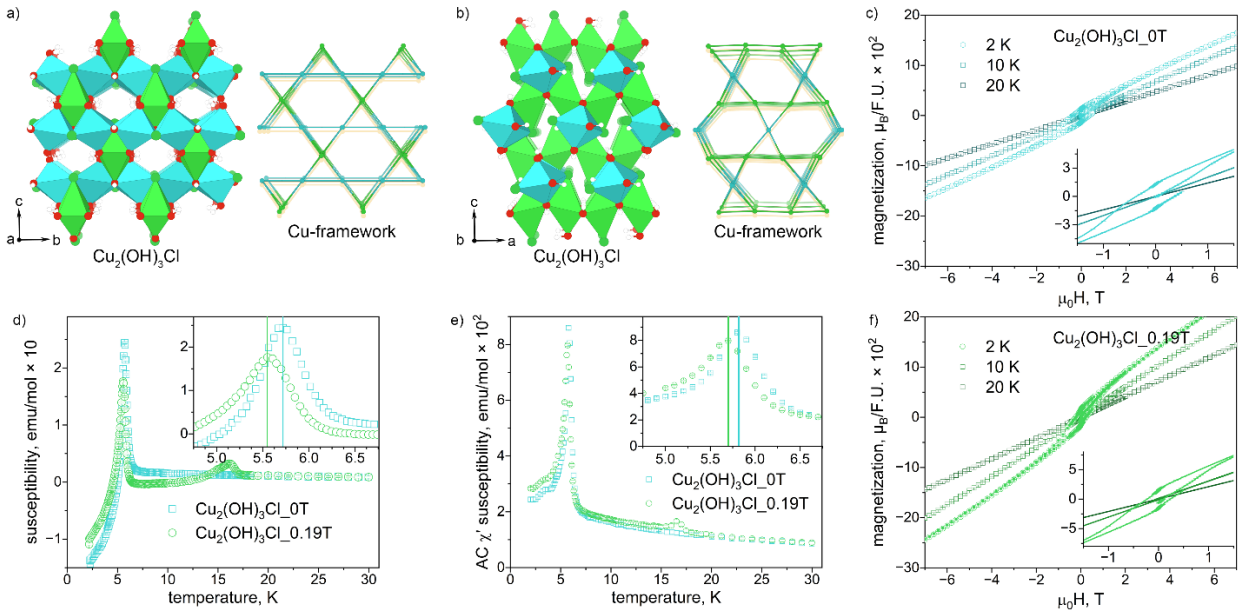

**Figure 8.** View of the $Cu_2(OH)_3Cl$ atacamite structure and Cu framework along (*a*) [001] and (*b*) [010] directions. White, red, dark green, and blue and green polyhedra represent H, O, and Cl atoms, and two sites for Cu octahedra, respectively. (*d*) ZFC DC magnetization and (*e*) AC magnetic susceptibility (2−30 K) for the $Cu_2(OH)_3Cl$ powder synthesized under various magnetic field strengths: 0 T (blue) and 0.19 T (green), respectively. Only real components ($\chi'$) are shown for AC susceptibility. Magnetization as a function of applied field at 2 K, 10 K and 20 K for (*c*) $Cu_2(OH)_3Cl\_0T$ and (*f*) $Cu_2(OH)_3Cl\_0.19T$ samples.

AC susceptibility and DC magnetization measurements (Figures 8d and 8e) on these two samples show a sharp peak at approximately $T_N \approx 5.7$ K, consistent with previous data on synthetic atacamite.[15,16] Intriguingly, the transition shifts by approximately 0.15 K between the 0 T and 0.19 T samples (Figures 8d and 8e). This is evident in both DC and AC χ' data, indicating that magnetosynthesis has affected the magnetic ground state. We also note the emergence of a peak in the DC magnetization at $T \approx 16.2$ K in the $Cu_2(OH)_3Cl\_0.19T$ sample, confirming the presence of a small amount of $Cu_3Cl_4(OH)_2 \cdot 2H_2O$. This may suggest that this phase, which we discussed above is not thermodynamically stable, is stabilized during this synthetic route by the presence of an applied magnetic field. Frequency-dependent AC susceptibility measurements performed with zero applied field on the $Cu_2(OH)_3Cl\_0T$ sample (Figure S27) display a shift in the $T_N$ to higher temperature with increasing frequency, confirming spin glass character of the magnetic ground state that was reported previously for synthetic atacamite.[15,16]

We performed Curie-Weiss fits from 100–350 K on DC magnetization data collected under an applied field of $\mu_0 H = 1$T (see Figure S28) and found $\Theta_{CW}$ of –46(2) and –70(3) K for the 0 T and 0.19 T samples, respectively, confirming the overall AFM nature of the magnetic interactions and the magnetic frustration of these samples (frustration index $f \approx 9$ and 14, respectively). This significant difference suggests that the application of a magnetic field during synthesis increased the magnetic frustration of atacamite.

The magnetization as a function of applied field at $T = 2$ K (Figures 8c and 8f) exhibit clear hysteresis loops with small net moments of 0.013 and 0.015 $\mu_B$/mol f.u. for the 0 T and 0.19 T samples, respectively. The coercive fields are approximately 0.25 and 0.38 T, respectively. Both samples display a small amount of wasp-waisted behavior, suggesting the coexistence of at least two competing interactions. Taken together, the MvsH behavior is consistent with canted AFM ground states in both samples, albeit with slight differences consistent with the temperature-dependent data.

DISCUSSION

As discussed above, this is one of the first studies on magnetosynthesis in 3$d$ systems, and the first on 3$d$ insulating system (recent work has been done on synthesizing metallic Co out of a Co-S flux).[44,45] The effects we see here seem much weaker than those observed in the (still few) 4$d$ and 5$d$ systems that have been studied so far.[7–9] This may be consistent with the hypothesis of Cao et al. that strong spin-orbit interactions can yield large structural and magnetic changes with magnetosynthesis,[7] although we note that the ~0.2-0.5% lattice changes we observe here are on the same order of magnitude reported for, e.g., BaIrO$_3$ (~0.7-0.85%).[8] This is therefore worthy of further theoretical study.

Overall, the only material out of the four studied here that definitively showed a shift in magnetic properties with magnetosynthesis is atacamite, which exhibits strong frustration and a slightly canted AFM ground state, by ~3%. While subtle, the differences between the 0 T and 0.19 T samples are robust. (Cu,Zn)$_3$Cl$_4$(OH)$_2$·2H$_2$O, which has a canted AFM ground state with a small net moment, exhibited changes in lattice parameters and coordination of Cu$^{2+}$ cations; we note that unfortunately we were not able to synthesize enough material under an applied field to compare magnetic properties. Moreover, the occupancy of Zn at the Cu1 site as a function of field is challenging to determine without growing crystals large enough for neutron diffraction yet can still affect magnetic properties.

Interestingly, the simple AFM (CuCl$_2$·2H$_2$O) and the QSL (HBS) did not exhibit a significant difference in magnetic properties ($T_N$ for CuCl$_2$·2H$_2$O and $\Theta_{CW}$ for both materials). While this was complicated by the broadness or lack of peaks in the susceptibility data and by the known variability in Curie-Weiss fits, we speculate that $CuCl_2 \cdot 2H_2O$'s AFM ground state is likely too stable for the energy scale of a small magnetic field (~0.009 meV per $S$=1/2 for the 0.09 T field to ~0.037 meV for the 0.37 T field) to perturb. We postulate that the degenerate frustrated AFM ground states—that also exhibit tiny net moments, likely due to spin canting—of $(Cu,Zn)_3Cl_4(OH)_2 \cdot 2H_2O$ and atacamite are accessible to these small energy scales. However, HBS, which is much more frustrated than the other materials, was not apparently affected; the AFM interactions are the strongest ($J \approx 15$ meV) out of all materials studied here, and HBS's spin gap (~1 meV) far exceeds the ~0.04 meV per $S$=1/2 scale of our magnetosynthesis.

We hypothesize that magnetosynthesis may affect the ground state in several ways: 1) it may induce a magnetostructural effect that influences the exchange interactions, which is borne out by the structural changes observed in $(Cu,Zn)_3Cl_4(OH)_2 \cdot 2H_2O$, and 2) it may "select" or stabilize a magnetic configuration with either competing interactions and/or a net moment (like we showed here with atacamite). Further study, likely involving developing novel computational techniques, will be needed to fully confirm these hypotheses and investigate how universal they are. The low-temperature methods that we develop here for incorporating a magnetic field during synthesis are novel; previous work placed permanent magnets outside a box furnace, severely limiting the strength that can be achieved. Thus, these methods can easily be applied to other materials families to continue exploring the use of magnetic field as a synthetic handle. In addition, detailed studies of transport properties, electron paramagnetic resonance spectroscopy, Raman vibrational modes, and photophysics can also help to unravel the effects of magnetosynthesis on many variables, including local structure, speciation of complexes, ion anisotropy, etc.

CONCLUSION

Magnetic field is a highly underexplored synthetic variable, and most work exploring its effect has been performed on 4$d$ and 5$d$ transition metal-containing compounds. Here, we performed the first systematic exploration of magnetosynthesis in 3$d$ compounds, focusing on materials containing $S=1/2$ $Cu^{2+}$ in a variety of lattices and with a range of magnetic frustration. We developed novel methods to easily incorporate a magnetic field into low-temperature hydrothermal and room-temperature evaporative synthesis techniques. We applied these methods to a series of materials that exhibit a range of low-temperature magnetic properties from QSL (HBS $Cu_3Zn(OH)_6Cl_2$) to simple low-temperature antiferromagnetism ($CuCl_2 \cdot 2H_2O$) to complex low-temperature antiferromagnetism exhibiting spin canting and/or spin-glass behavior ($Cu_3Cl_4(OH)_2 \cdot 2H_2O$) and atacamite $Cu_2(OH)_3Cl$.

Notably, we observed the stabilization of an understudied phase $Cu_3Cl_4(OH)_2 \cdot 2H_2O$ with a small amount of Zn (Cu:Zn 2.85:0.15) and report the first single crystal structural determination of this phase as well as its magnetic properties. Its Cu/Zn lattice consists of a stretched and highly distorted kagome arrangement. Synthesis of this phase under a 0.19 T magnetic field resulted in subtle distortion of the unit cell. Most intriguingly, atacamite $Cu_2(OH)_3Cl$ exhibited a 0.15 K (~3%) decrease of its $T_N$ with magnetosynthesis under a 0.19 T field. The Curie-Weiss temperature $\Theta_{CW}$ also became more negative, indicating that the AFM correlations grew stronger and the material became more magnetically frustrated. The results suggest that magnetosynthesis can affect the structure and magnetic ground state of frustrated materials with 3$d$ transition metals.

**Supporting Information**. The following files are available free of charge at ??.

Additional synthetic details; scanning electron microscopy data; crystallographic details of the SCXRD structures; PXRD data and crystallographic data; and additional magnetism data (PDF)

**Accession codes**.

Deposition numbers 2499485–2499487 contain the supplementary crystallographic data for this paper. These data can be obtained free of charge via the joint Cambridge Crystallographic Data Centre (CCDC) and Fachinformationszentrum Karlsruhe Access Structures service.

AUTHOR INFORMATION

**Corresponding Authors**

* Anna.Berseneva@NREL.gov; Rebecca.Smaha@NREL.gov

**Present Addresses**

†Materials Science & Engineering, University of Colorado Boulder, Boulder, Colorado USA

**Author Contributions**

The manuscript was written through contributions of all authors. All authors have given approval to the final version of the manuscript. M.P and A.B. contributed to manuscript equally.

ACKNOWLEDGMENT

This work was authored by NREL for the U.S. Department of Energy (DOE), operated under Contract No. DE-AC36-08GO28308. M.P. (synthesis, characterization) was supported by the U.S. Department of Energy, Office of Science, Office of Workforce Development for Teachers and Scientists (WDTS) under the Science Undergraduate Laboratory Internships (SULI) program. A.B. (characterization, supervision) was supported by the Director's Fellowship within the Laboratory Directed Research and Development (LDRD) Program at NREL. Funding for supervision and calculations was provided by the U.S. Department of Energy, Office of Science, Basic Energy Sciences, Division of Materials Science, through the Office of Science Funding

Opportunity Announcement (FOA) No. DE-FOA-0002676: Chemical and Materials Sciences to Advance Clean-Energy Technologies and Transform Manufacturing. This research used resources of the Center for Functional Nanomaterials, which is a US Department of Energy (DOE) Office of Science Facility, and the Scientific Data and Computing Center, a component of the Computational Science Initiative, at Brookhaven National Laboratory under contract DE-SC0012704. The authors thank Y.S. Lee and J.R. Neilson for helpful discussions. The views expressed in the article do not necessarily represent the views of the DOE or the U.S. Government.REFERENCES

(1) Ramirez, A. P. Strongly Geometrically Frustrated Magnets. *Annual Review of Materials Research* **1994**, *24*, 453–480. https://doi.org/10.1146/annurev.ms.24.080194.002321.
(2) Semeghini, G.; Levine, H.; Keesling, A.; Ebadi, S.; Wang, T. T.; Bluvstein, D.; Verresen, R.; Pichler, H.; Kalinowski, M.; Samajdar, R.; Omran, A.; Sachdev, S.; Vishwanath, A.; Greiner, M.; Vuletić, V.; Lukin, M. D. Probing Topological Spin Liquids on a Programmable Quantum Simulator. *Science* **2021**, *374* (6572), 1242–1247. https://doi.org/10.1126/science.abi8794.
(3) Smaha, R. W.; He, W.; Sheckelton, J. P.; Wen, J.; Lee, Y. S. Synthesis-Dependent Properties of Barlowite and Zn-Substituted Barlowite. *Journal of Solid State Chemistry* **2018**, *268*, 123–129. https://doi.org/10.1016/j.jssc.2018.08.016.
(4) Smaha, R. W.; He, W.; Jiang, J. M.; Wen, J.; Jiang, Y.-F.; Sheckelton, J. P.; Titus, C. J.; Wang, S. G.; Chen, Y.-S.; Teat, S. J.; Aczel, A. A.; Zhao, Y.; Xu, G.; Lynn, J. W.; Jiang, H.-C.; Lee, Y. S. Materializing Rival Ground States in the Barlowite Family of Kagome Magnets: Quantum Spin Liquid, Spin Ordered, and Valence Bond Crystal States. *npj Quantum Mater.* **2020**, *5* (1), 23. https://doi.org/10.1038/s41535-020-0222-8.
(5) Rivoirard, S. High Steady Magnetic Field Processing of Functional Magnetic Materials. *JOM* **2013**, *65* (7), 901–909. https://doi.org/10.1007/s11837-013-0619-y.
(6) Guillon, O.; Elsässer, C.; Gutfleisch, O.; Janek, J.; Korte-Kerzel, S.; Raabe, D.; Volkert, C. A. Manipulation of Matter by Electric and Magnetic Fields: Toward Novel Synthesis and Processing Routes of Inorganic Materials. *Materials Today* **2018**, *21* (5), 527–536. https://doi.org/10.1016/j.mattod.2018.03.026.
(7) Cao, G.; Zhao, H.; Hu, B.; Pellatz, N.; Reznik, D.; Schlottmann, P.; Kimchi, I. Quest for Quantum States via Field-Altering Technology. *npj Quantum Mater.* **2020**, *5* (1), 83. https://doi.org/10.1038/s41535-020-00286-2.
(8) Cao, T. R.; Zhao, H.; Huai, X.; Quane, A.; Tran, T. T.; Ye, F.; Cao, G. Field-Tailoring Quantum Materials: Magneto-Synthesis of Metastable Metallic Phases in a Trimer Iridate. arXiv August 18, 2025. https://doi.org/10.48550/arXiv.2508.07545.
(9) Pellatz, N.; Kim, J.; Kim, J.-W.; Kimchi, I.; Cao, G.; Reznik, D. Magnetosynthesis Effect on Magnetic Order, Phonons, and Magnons in Single-Crystal $Sr_2IrO_4$. *Phys. Rev. Materials* **2023**, *7* (12), 123802. https://doi.org/10.1103/PhysRevMaterials.7.123802.27